\documentclass[10pt,sigconf,authorversion,letterpaper]{acmart}

\usepackage[inline]{enumitem}
\usepackage{subfig}
\usepackage{graphicx}
\usepackage{todonotes}

\AtBeginDocument{%
  \providecommand\BibTeX{{%
    \normalfont B\kern-0.5em{\scshape i\kern-0.25em b}\kern-0.8em\TeX}}}

\begin{document}

\copyrightyear{2021}
\acmYear{2021}
\setcopyright{acmlicensed}\acmConference[ANRW '21]{Applied Networking Research Workshop}{July 24--30, 2021}{Virtual Event, USA}
\acmBooktitle{Applied Networking Research Workshop (ANRW '21), July 24--30, 2021, Virtual Event, USA}
\acmPrice{15.00}
\acmDOI{10.1145/3472305.3472316}
\acmISBN{978-1-4503-8618-0/21/07}

\title{Adaptive Cheapest Path First Scheduling in a Transport-Layer Multi-Path Tunnel Context}

\author{Marcus Pieska}
\orcid{1234-5678-9012}
\author{Alexander Rabitsch}
\email{firstname.lastname@kau.se}
\affiliation{%
  \institution{Computer Science Department}
  \streetaddress{P.O. Box 1212}
  \city{Karlstads Universitet}
  \state{Karlstad}
  \country{Sweden}
  \postcode{43017-6221}
}

\author{Anna Brunstrom}
\author{Andreas Kassler}
\email{firstname.lastname@kau.se}
\affiliation{%
  \institution{Computer Science Department}
  \streetaddress{P.O. Box 1212}
  \city{Karlstads Universitet}
  \state{Karlstad}
  \country{Sweden}
  \postcode{43017-6221}
}

\author{Markus Amend}
\email{markus.amend@telekom.de}
\affiliation{%
  \institution{Deutsche Telekom}
  \streetaddress{Deutsche-Telekom-Allee 9}
  \city{Darmstadt}
  \country{Germany}}

\renewcommand{\shortauthors}{Pieska, et al.}


\begin{abstract}
Bundling multiple access technologies increases capacity, resiliency and robustness of network connections. Multi-access is currently being standardized in the ATSSS framework in 3GPP, supporting different access bundling strategies. Within ATSSS, a multipath scheduler needs to decide which path to use for each user packet based on path characteristics. The \textit{Cheapest Path First} (CPF) scheduler aims to utilize the cheapest path (e.g. WiFi) before sending packets over other paths (e.g. cellular). In this paper, we demonstrate that using CPF with an MP-DCCP tunnel may lead to sub-optimal performance. This is due to adverse interactions between the scheduler and end-to-end and tunnel congestion control. Hence, we design the Adaptive Cheapest Path First (\textit{ACPF}) scheduler that limits queue buildup in the primary bottleneck and moves traffic to the secondary path earlier. We implement ACPF over both TCP and DCCP congestion controlled tunnels. Our evaluation shows that ACPF improves the average throughput over CPF between 24\% to 86\%.  
\end{abstract}




\maketitle

\section{Introduction}
Fifth generation cellular networks (5G) aim to support multi-access communication by using multiple wireless technologies through \textit{Access Traffic Steering, Switching, and Splitting} (ATSSS, see 3GPP Rel. 16 \cite{3gpp.23.501}). Efficient bundling under a common framework can increase the capacity, resiliency and robustness of wireless networks. ATSSS supports flexible traffic steering and transparent capacity aggregation of multiple access networks. In this architecture, end-to-end TCP connections can be split at an anchor, transformed into \textit{Multipath TCP} (MPTCP) \cite{mptcpv1} and routed over multiple wireless access networks between anchor and user. Unreliable traffic can instead be sent over \textit{multi-path tunnels} \cite{3gpp.23.700} using novel multipath protocols that do not enforce strict reliability, e.g. \textit{Multipath DCCP} (MP-DCCP) \cite{mpdccp} or \textit{Multipath QUIC} (MP-QUIC) \cite{mpquic} with the datagram extension \cite{quic-datagram}.

Transparent and flexible \textit{path aggregation} and prioritization of different access technologies is an important goal for ATSSS. Aggregation requires a packet scheduler, which determines the path for each packet. Ideally, such a scheduler should fully aggregate the available capacity of all access networks, which is difficult due to different path characteristics such as latency or available capacity. As ATSSS aims to offer cellular users aggregation of \textit{non-trusted} wireless access, schedulers that express a preference for one or the other become desirable. Many cellular users would benefit from off-loading traffic onto WiFi whenever possible, and thereby save their limited cellular tariff --- a trade-off which is also attractive to telecom operators. A scheduler may prioritize paths based on different criteria \cite{iccrg-schedulers}, including delay or \textit{economic} aspects like \textit{cost}.  In that context, the \textit{strict priority} scheduler \cite{iccrg-schedulers} aims to send packets over the primary path until the capacity is exhausted. 

In this paper we: 
\begin{enumerate}[label=(\roman*)]  
\item show that using CPF in a multipath congestion controlled tunneling framework leads to sub-optimal performance both when using DCCP and TCP-based tunnels. The main reason is the detrimental interaction between the scheduler, the per path congestion control algorithms and the end-host congestion controller.   
\item design and implement the novel \textit{Adaptive Cheapest Path First} (ACPF) scheduler, which uses the secondary path earlier, when ACPF infers congestion on the primary path. ACPF effectively limits congestion in favor of utilizing the secondary path, when the secondary path is underutilized. 
\item implement and integrate ACPF with BBR-like congestion control over individual tunnels in both the Linux kernel and a user-space framework. 
\item evaluate ACPF and show that it copes more effectively with the presence of nested congestion control loops, and that ACPF solicit the end-to-end connection to maintains more packets in-flight.
\end{enumerate}

The rest of the paper is structured as follows: Section \ref{sec:rel-main} reviews related work. Section \ref{sec:des-main} illustrates why CPF behaves poorly when used with congestion controlled tunnels and presents the design of our new ACPF scheduler. Section \ref{sec:imp-main} details the implementation of ACPF. Section \ref{sec:res-main} compares the performance of CPF and ACPF. The paper concludes in section \ref{sec:end-main} discussing future work. 

\begin{figure*}[ht]
\subfloat[][Basic ACPF design; note the tunnel \& path terminology.]{
\includegraphics[width=.49\linewidth]
{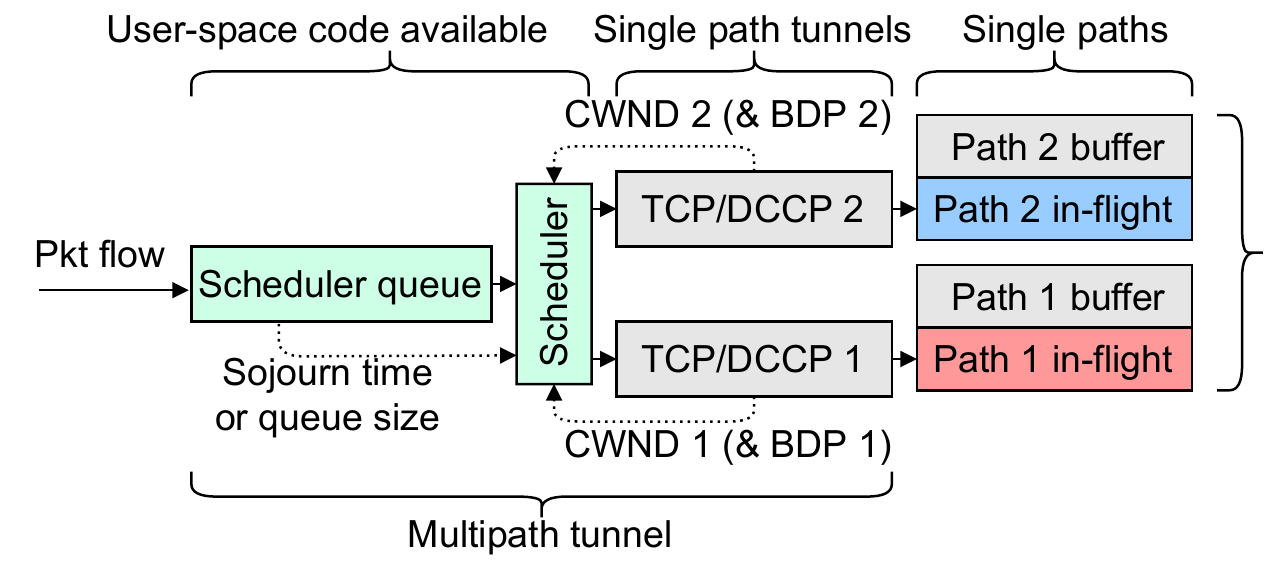}
\label{fig:ppp-design}}%
\subfloat[][CPF issue to the left \& ACPF solution to the right]{
\hspace{15mm}\includegraphics[width=.33\linewidth]
{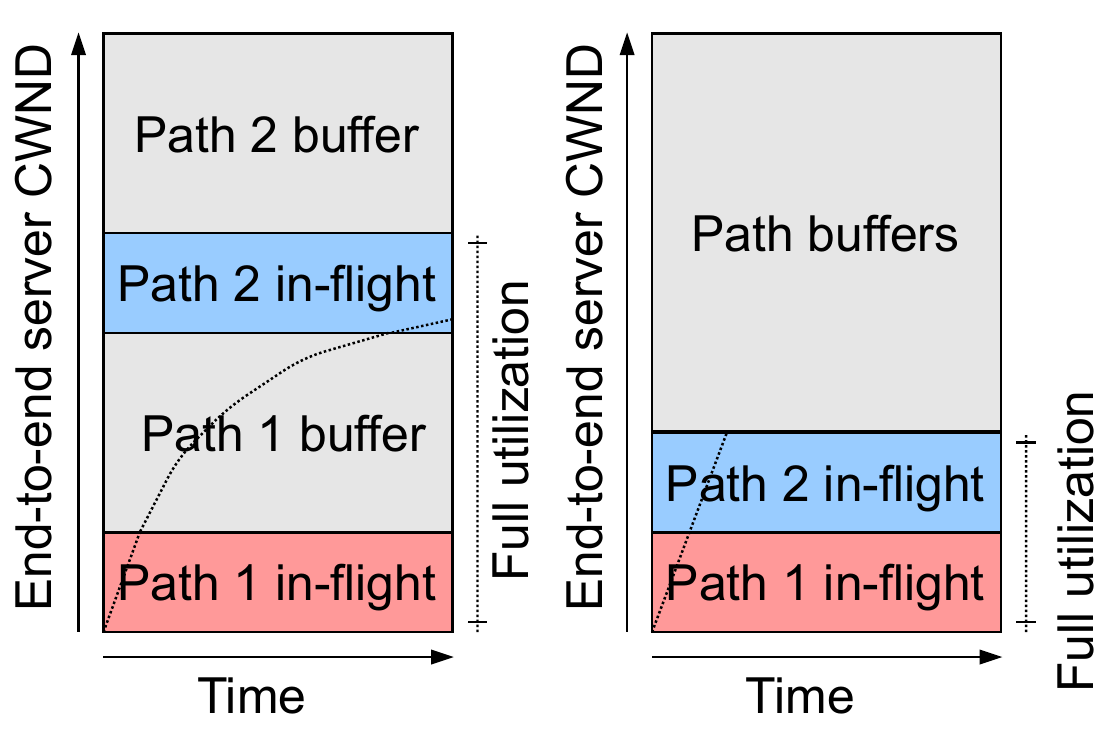}\hspace{15mm}
\label{fig:ppp-problem}}

\caption{\textbf{Fig. a)} High level ACPF illustration. Green modules are available as both user-space code and kernel code. \textbf{Fig.  b)} Server (end-to-end) $CWND_s$ growth, and where over each path the $CWND_s$ is distributed over time.}
\label{fig:ppp-design-problem}
\end{figure*}

\section{Related Work}
\label{sec:rel-main}
  
Packet scheduling over multiple paths is an active area of research \cite{iccrg-schedulers, peekaboo, mptcp-ll-sched}. An early evaluation of MPTCP schedulers \cite{paasch2014} revealed that
\begin{enumerate*}[label=(\roman*)]
\item \textit{head-of-line blocking} and
\item \textit{receive window blocking} reduce the performance, in particular when the delay over the paths is highly asymmetrical. \end{enumerate*} The MinRTT scheduler \cite{minRTT} is the default MPTCP scheduler, and prioritizes the path with the shortest estimated RTT. Blocking Estimation-based MPTCP scheduler algorithm \cite{ferlin2016} improves upon MinRTT by better estimating the amount of in-flight packets over each path. Other algorithms include Earliest Completion First \cite{ECF}, Peekaboo \cite{peekaboo}, and DEMS \cite{guo2017}. 

\textit{Congestion control algorithms} (CCA) tend to regulate the number of in-flight packets to be no more than the congestion window ($CWND$). TCP-NewReno \cite{floyd1999newreno} is an early CCA which increases $CWND$ with one packet per RTT until a loss is detected, at which point $CWND$ is reduced. TCP-Cubic \cite{tcpCubic} adjusts $CWND$ according to a cubic function, but reacts to loss much like NewReno. More recently, TCP-BBR \cite{cardwell2016bbr} tracks the throughput and the minimum RTT. BBR may then utilize the path fully by having $CWND$ equal roughly two bandwidth-delay products (BDP). BBR has been implemented for DCCP under the name CCID5\footnote{\url{https://github.com/telekom/mp-dccp/tree/d9b4b7471a69184105885abbc4d45c011b82543d/net/dccp/ccids}}. Estimating the BDP when using a CCA other than BBR requires tracking the minimum $RTT_{min}$ and limiting $CWND$ such that the $RTT$ does not exceed $RTT_{min}$ \cite{mpt-bm}. Finding a CCA suitable for wireless is an ongoing effort \cite{haile2021end}.

Explicit congestion notification (ECN) \cite{welzl2010congestion} is a method whereby the \textit{network} indicates the presence of congestion to the sending-side transport layer --- \textit{as opposed to have the sender infer the amount of congestion}. CoDel \cite{codel} aims to control congestion by dropping packets when a certain queue latency is reached, thus triggering the end-to-end congestion control to reduce its sending rate.

\section{Design, Issues, \& Solutions}
\label{sec:des-main}

\begin{figure}

\centering
\subfloat[][Topology; TBF sets throughput; NETEM sets delay.]{
\includegraphics[width=1.0\linewidth]
{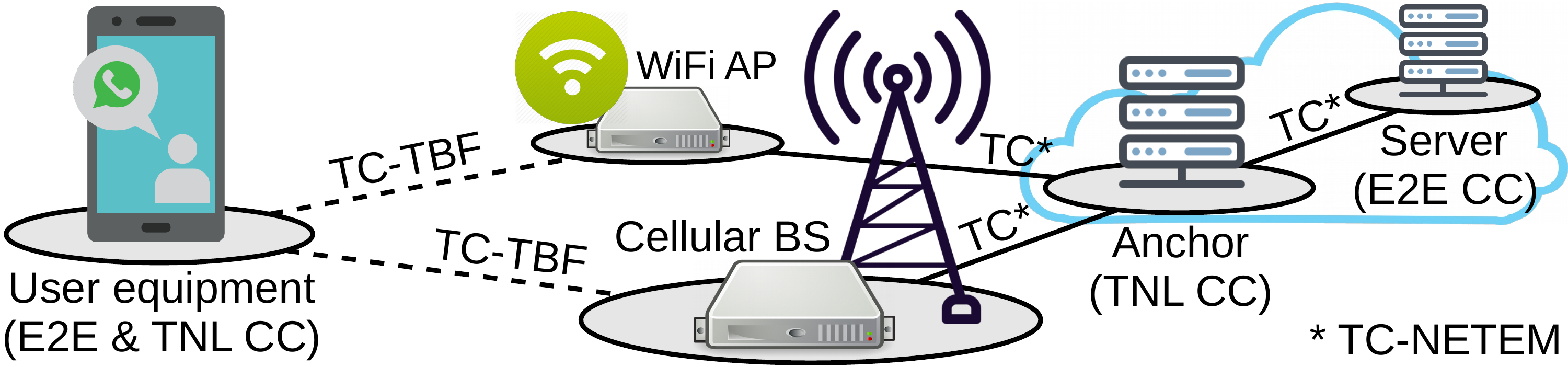}
\label{fig:topology}} \par

\centering
\subfloat[][Pseudo code for generic ACPF management.]{
\includegraphics[width=.75\linewidth]
{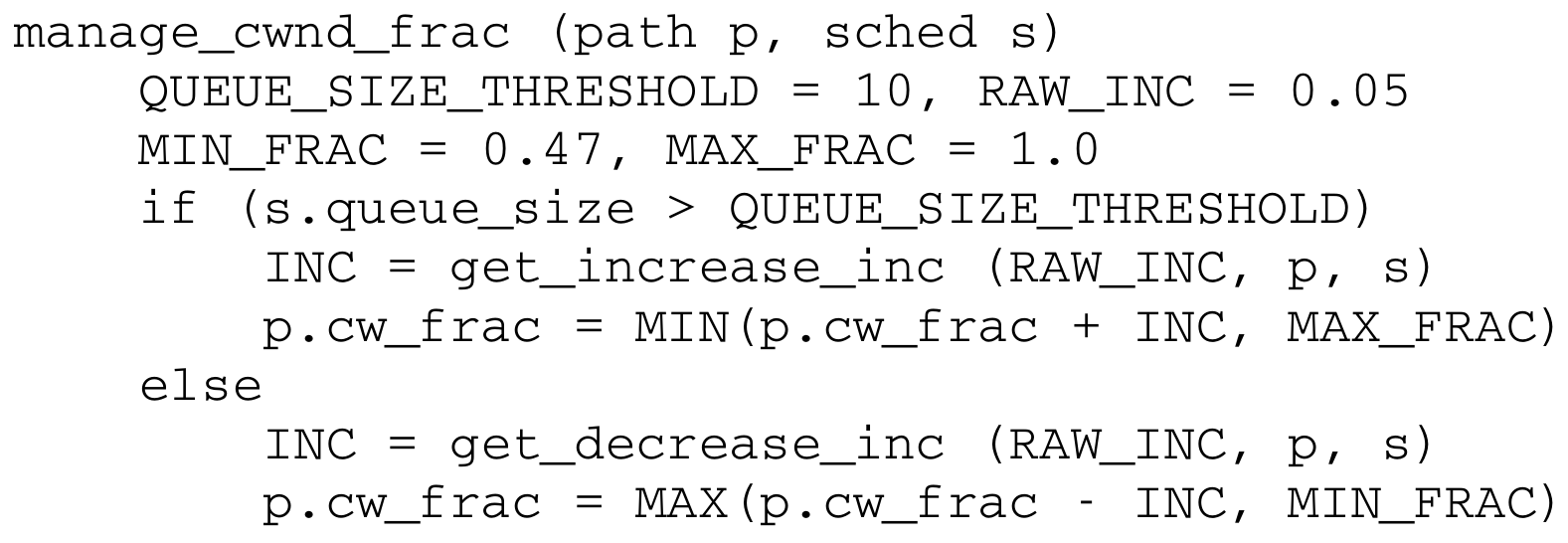}
\label{fig:ppp-mng-func}} \par

\centering
\subfloat[][Pseudo code for user-space ACPF management.]{
\includegraphics[width=.75\linewidth]
{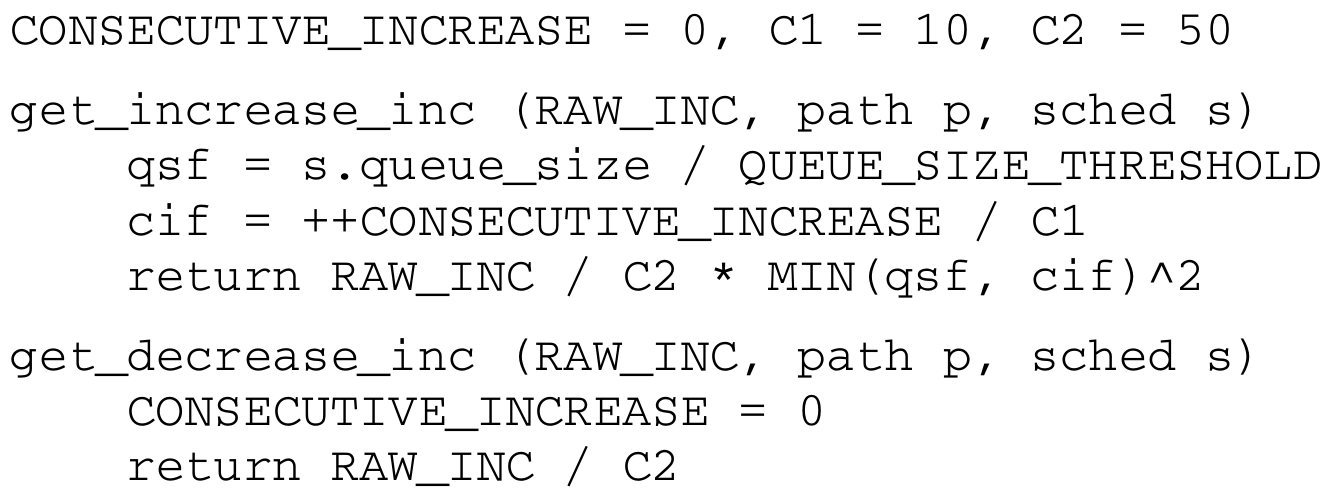}
\label{fig:ppp-mng-func-us}} 

\caption{Multi-path context \& ACPF logic.}
\end{figure}

In the multi-access context, we assume $p$ paths --- \textit{and single-path tunnels} --- between the UE and the anchor, with each path having an independent congestion control regulating $CWND_p$ --- \textit{per Figure \ref{fig:topology}}. Packets enter the multi-access framework via a scheduler queue, and for each packet a scheduler decides over which path to send it. At the anchor, packets leave the framework and are forwarded to the server. Thus, client and server applications need not to be modified to use the multi-access solution. Next, we review CPF and its performance problem in the multi-access context and illustrate the design rationale for ACPF.

\subsection{Cheapest Path First}
\label{sec:des-cpf}

The \textit{Cheapest Path First} (CPF, or \textit{strict priority} \cite{iccrg-schedulers}) packet scheduler ranks available paths according to cost in ascending order --- \textit{i.e., the cheapest path having highest priority, e.g. WiFi first, then Cellular}. The scheduler will consider path $p$ \textit{fully utilized} if the number of in-flight packets --- \textit{or bytes} --- is equal to or greater than $CWND_p$. If this condition holds, path $p$ will be marked as \textit{not available} and the scheduler will proceed to schedule packets over lower priority path $p+1$. 

The problem with CPF is illustrated in Figure \ref{fig:ppp-problem}, which shows how the server $CWND_s$ grows, and which parts of the multipath tunnel it occupies over time. The ideal outcome is to the right, whereas the outcome to the left is what tends to happen --- \textit{as shown in section \ref{sec:res-main}}. As $CWND_s$ grows, $CWND_1$ tends to grow in parallel, leading CPF to schedule mostly on the primary path. This parallel growth can \textit{delay}, or even \textit{preclude} aggregation, depending on the server CCA. Most similar to Figure \ref{fig:ppp-problem} is when the server uses NewReno; letting $CWND_s$ into the path 1 bottleneck buffer will \textit{elongate} path 1, and slow down the $CWND_s$ growth. This is seen in Figure~\ref{fig:ppp-problem} as a decreasing growth rate of $CWND_s$ to the left. Ideally, the \textit{size} of $CWND_1$ is such that $CWND_s$ first occupies path 1, then path 2, \textit{and lastly} the the bottleneck buffers. This is our goal with ACPF, as seen to the right.

\subsection{Adaptive Cheapest Path First}
\label{sec:des-acpf}

The main problem with CPF is that packets are sent over the cheaper path when they \textit{should} be sent over the secondary path. The former will \textit{reduce throughput} and \textit{increase latency}, while the opposite is true for the latter. CoDel and ECN could be used at the bottleneck and would shift packets onto the secondary path by \textit{shrinking} $CWND_1$. The main idea of ACPF is to use the secondary path \textit{earlier}, when \textit{congestion can be inferred} on the primary path and the sender still wants to increase its sending rate. The goal is to avoid occupying bottleneck buffers until the in-flight capacity (i.e. the BDP) have been reached on all paths. To achieve that, the amount of packets that can be scheduled over any path $p$ is limited to a fraction of $CWND_p$, ideally \textit{corresponding to the BDP} of path $p$. We denote $CWND_p$ as \textit{raw}, and the fraction as the \textit{live $CWND_p$}. The fraction is \textit{periodically decreased} if the number of packets in the scheduling queue is below a given limit, and \textit{vice versa}. This results in an \textit{initial shallow occupancy} of the primary path buffer along with an earlier utilization of the secondary path. This is \textit{not} congestion control, as the congestion reduction is likely often temporary. This key management is illustrated in Figure \ref{fig:ppp-design}, with the inputs being the scheduling queue occupancy and $CWND_p$.

Using TCP-BBR (or DCCP-CCID5) as tunnel CCA is particularly helpful to ACPF. BBR periodically drains the path to estimate the minimum RTT ($RTT_{min_p}$), while it will continuously probe for the throughput ($BW_p$). It will then set  $CWND_p$ according to equation \ref{eq:bbr-cwnd} where $G$ is a gain typically set to 2. Thus, since the BBR $CWND$ is based on the BDP, ACPF may adjust how deeply the bottleneck buffer is filled by adjusting the live $CWND_p$ in equation \ref{eq:bbr-cwnd} with a factor $AG$. Since the gain $G$ is 2, setting $AG$ to 0.5 will reduce the bottleneck buffer occupancy to near zero, while a factor of 1 will allow for the standard BBR buffer occupancy of one BDP. A important feature of BBR is that we know \textit{approximately} what the \textit{minimum gain} should be to achieve aggregation \textit{without} needlessly splitting flows. Without this, ACPF would have no lower limit on $AG$, and if the live $CWND_p$ were to become smaller than the BDP \textit{the flow would be split needlessly}. Figure \ref{fig:ppp-mng-func} contains pseudo code which illustrates how to manage $AG$ when working with BBR by \textit{inferring} that the BDP should be roughly half the $CWND_p$.     

\begin{equation}
CWND_p=RTT_{min_p}*BW_p*G\label{eq:bbr-cwnd}
\end{equation}

\section{Implementation}
\label{sec:imp-main}

We implemented ACPF \textit{on top} of CPF as a management function \textit{based on} Figure \ref{fig:ppp-mng-func}. Each path has its own $CWND_p$ fraction, which is kept synchronized. A \textit{queue size threshold} of 10 packets is used to determine the direction of the $CWND_p$ fraction adjustment; a queue size above will cause the live $CWND_p$ to grow and vice versa. The minimum fraction was determined \textit{experimentally} to be 0.47. ACPF has been implemented in the Open Source MP-DCCP Linux Kernel reference implementation\footnote{\url{https://multipath-dccp.org}}, as well as in a \textit{protocol agnostic} user-space framework based on MP-DCCP.

\subsection{Kernel-space ACPF}
\label{sec:imp-kern}

Kernel ACPF calls the $CWND_p$ management function (Figure \ref{fig:ppp-mng-func}) to update the $CWND_p$ fractions on \textit{all} paths, whenever a packet is scheduled onto any path. The rate at which the live $CWND_p$ fraction is adjusted per packet (\texttt{get\_increase\_inc} and \texttt{get\_decrease\_inc} in Figure \ref{fig:ppp-mng-func}) is set according to equation \ref{eq:acpf-kernel-inc}. This allows the $CWND_p$ fraction to increase as needed even for small $CWND_p$, while remaining stable for larger $CWND_p$. When BBR enters the \textit{ProbeRTT} state it is likely that the scheduler buffer occupancy temporarily increases as the server still sends at high rate. As the rate at which the $CWND_p$ fraction is adjusted is a function of the $CWND_p$, this can lead to an unnecessarily rapid growth of the $CWND_p$ fractions. The $CWND_p$ management is therefore temporarily disabled whenever BBR is in this state.

\begin{equation}
INC=RAW\_INC / CWND_p\label{eq:acpf-kernel-inc}
\end{equation}

\subsection{User-space ACPF}
\label{sec:imp-user}

User-space ACPF will invoke its $CWND_p$ management function in a dedicated thread every 1 ms. Using the naming convention in Figure \ref{fig:ppp-mng-func}, Figure \ref{fig:ppp-mng-func-us} illustrates how the increase rate is scaled. The increment $INC$ is scaled down by a factor $C_2=50$, leading to a \textit{default} time required to bring the fraction from its minimum to its maximum of 530 ms. However, ACPF is designed to react more aggressively to a queue-buildup which is \textit{persistent}, as opposed to transient. Given the 1 ms invocation interval, $C_1=10$ determines that any queue lasting for more than 10 ms is persistent. Without $C_1$ and $C_2$, the management becomes too aggressive.

\section{Evaluation}
\label{sec:res-main}

\begin{figure}
\centering

\subfloat[][CPF, short RTT, E2E CCA (from top): NewReno; Cubic; BBR.]{
\includegraphics[width=.98\linewidth]
{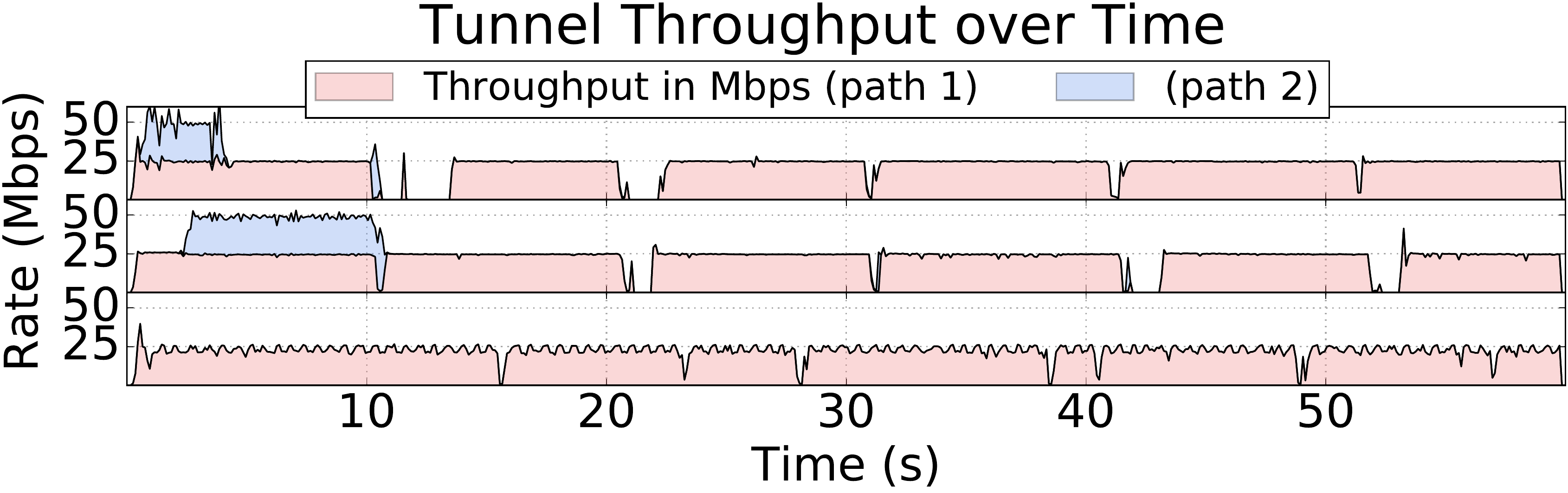} 
\label{fig:ts-comp-kernel-cpf-short}} \par

\subfloat[][ACPF, short RTT, E2E CCA (from top): NewReno; Cubic; BBR.]{
\includegraphics[width=.98\linewidth]
{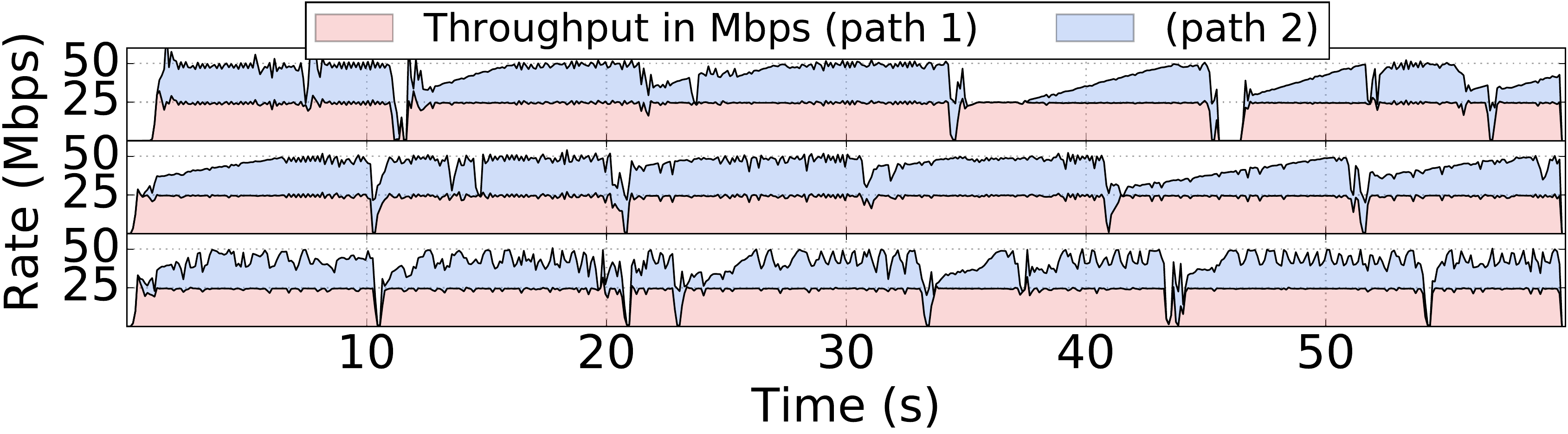} 
\label{fig:ts-comp-kernel-acpf-short}} \par

\subfloat[][CPF, long RTT, E2E CCA (from top): NewReno; Cubic; BBR.]{
\includegraphics[width=.98\linewidth]
{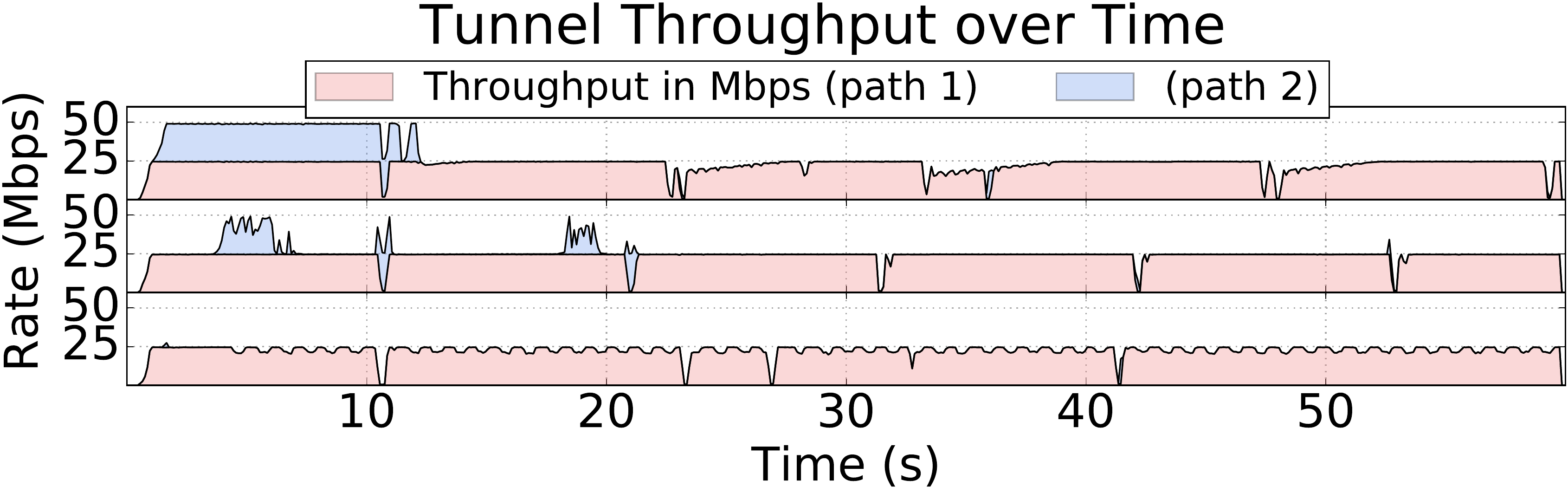} 
\label{fig:ts-comp-kernel-cpf-long}} \par

\subfloat[][ACPF, long RTT, E2E CCA (from top): NewReno; Cubic; BBR.]{
\includegraphics[width=.98\linewidth]
{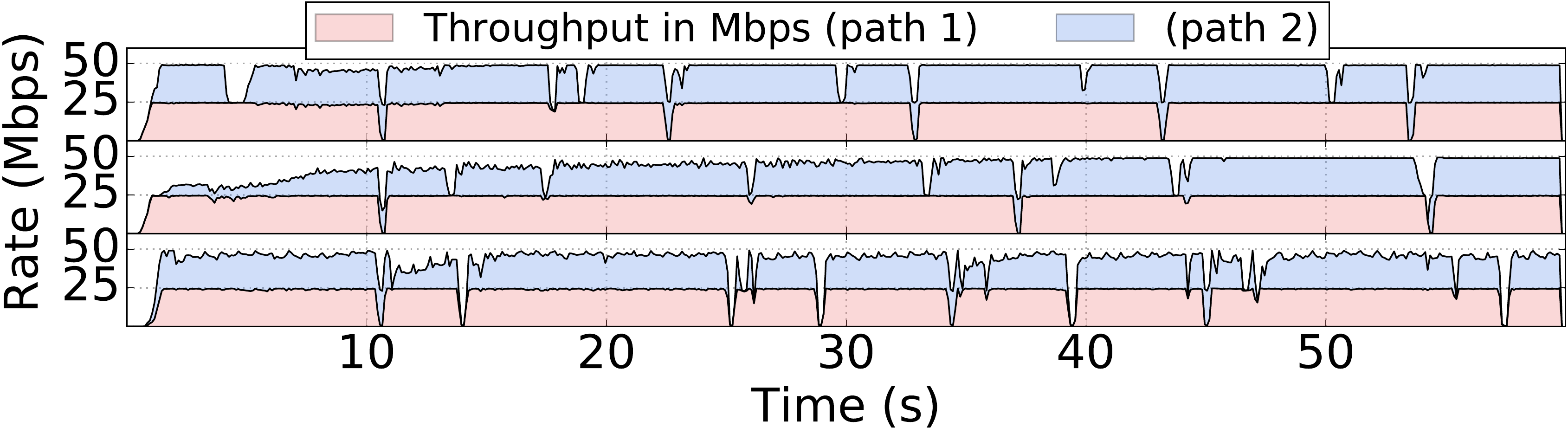}
\label{fig:ts-comp-kernel-acpf-long}} 

\caption{Kernel-space comparison of CPF \& ACPF.}
\end{figure}

\begin{figure*}
\centering

\subfloat[][CPF, short RTT, E2E CCA: TCP-NewReno.]{
\includegraphics[width=.48\linewidth]
{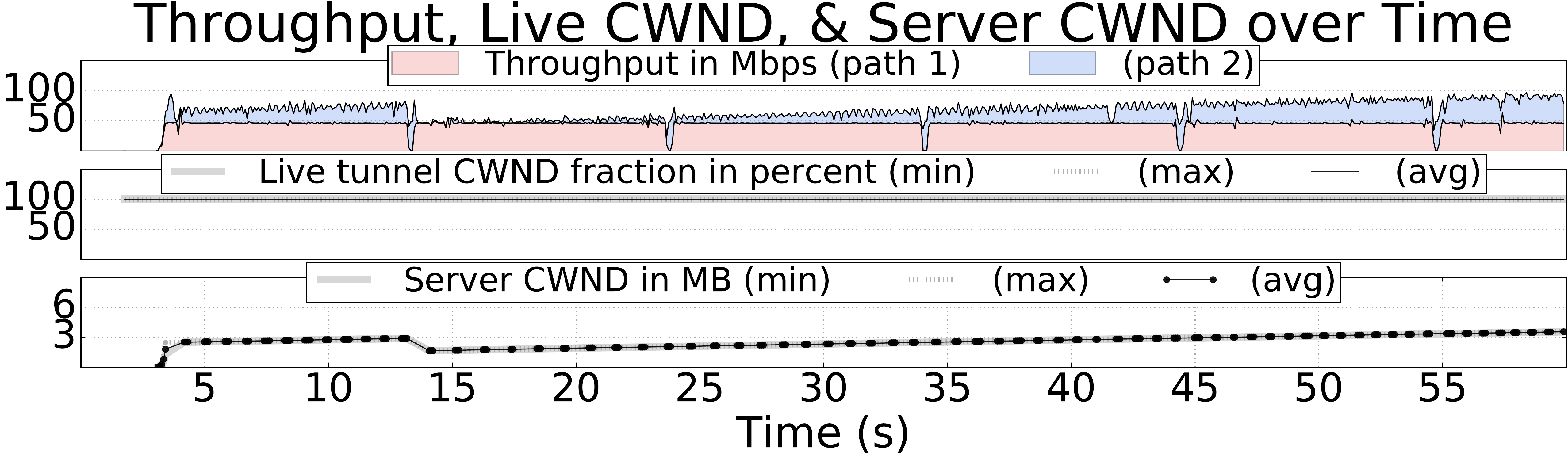}
\label{fig:ts-reno-bbr-pp}}
\subfloat[][ACPF, E2E CCA: TCP-NewReno.]{
\includegraphics[width=.48\linewidth]
{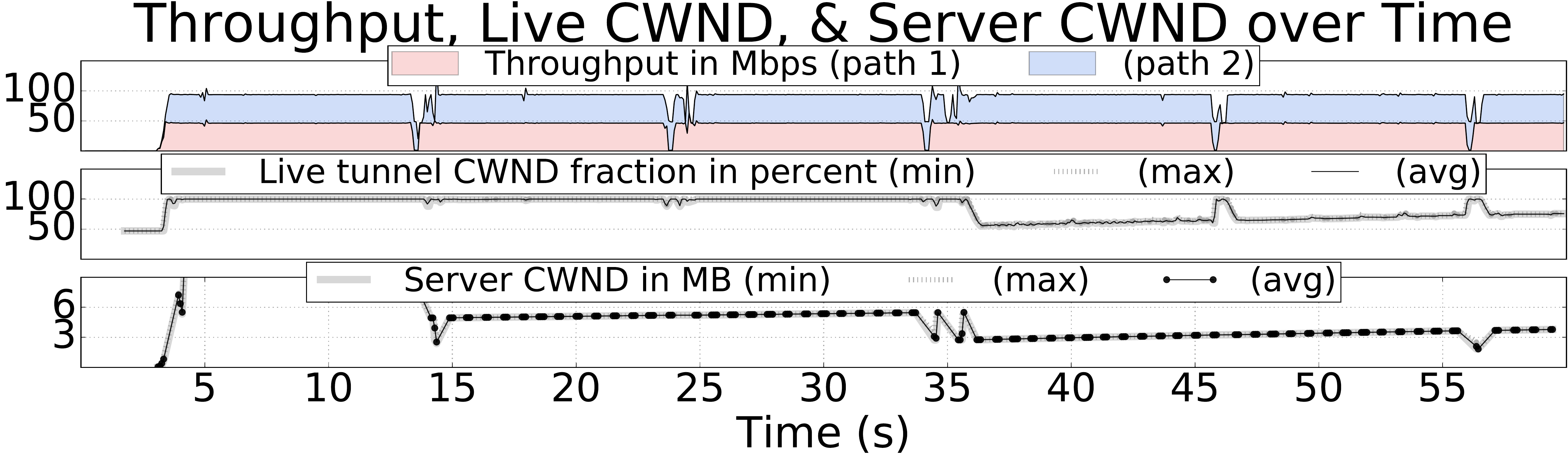}
\label{fig:ts-reno-bbr-ppp}} \par

\subfloat[][CPF, short RTT, E2E CCA: TCP-Cubic.]{
\includegraphics[width=.48\linewidth]
{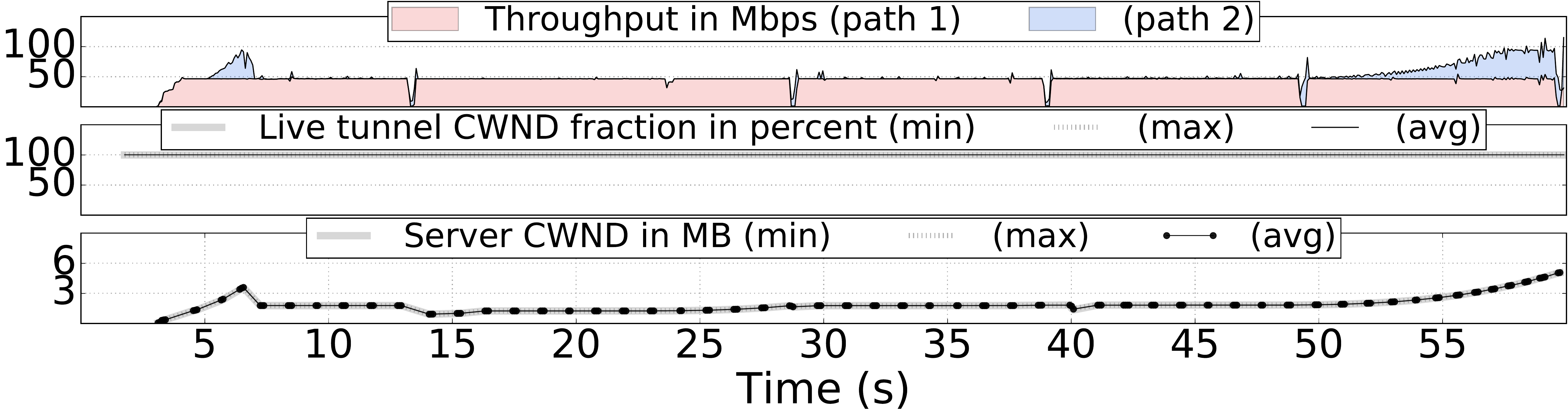}
\label{fig:ts-cubic-bbr-pp}}
\subfloat[][ACPF, short RTT, E2E CCA: TCP-Cubic.]{
\includegraphics[width=.48\linewidth]
{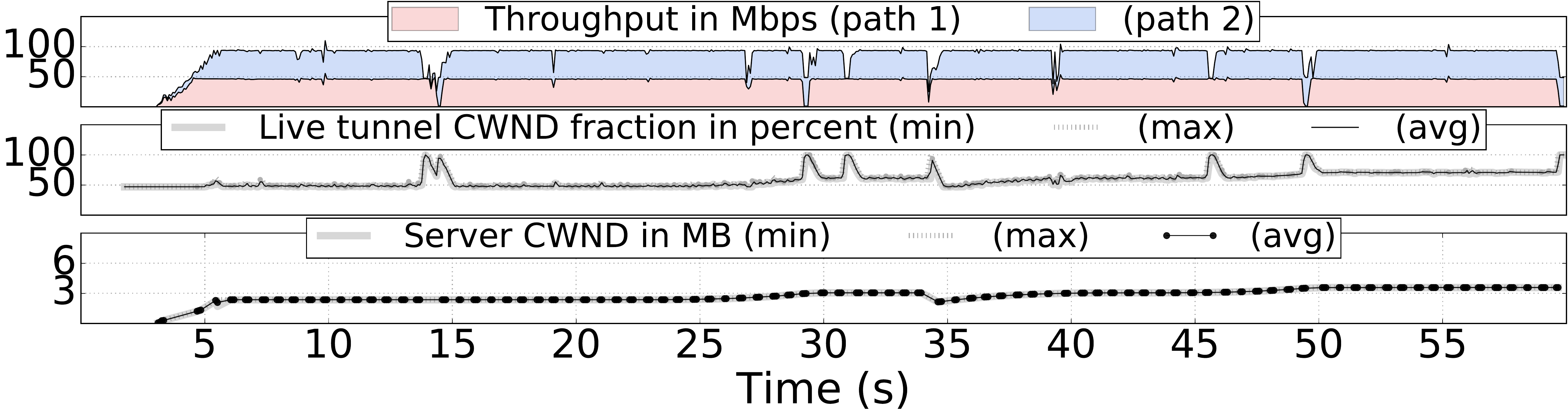}
\label{fig:ts-cubic-bbr-ppp}} \par

\subfloat[][CPF, short RTT, E2E CCA: TCP-BBR.]{
\includegraphics[width=.48\linewidth]
{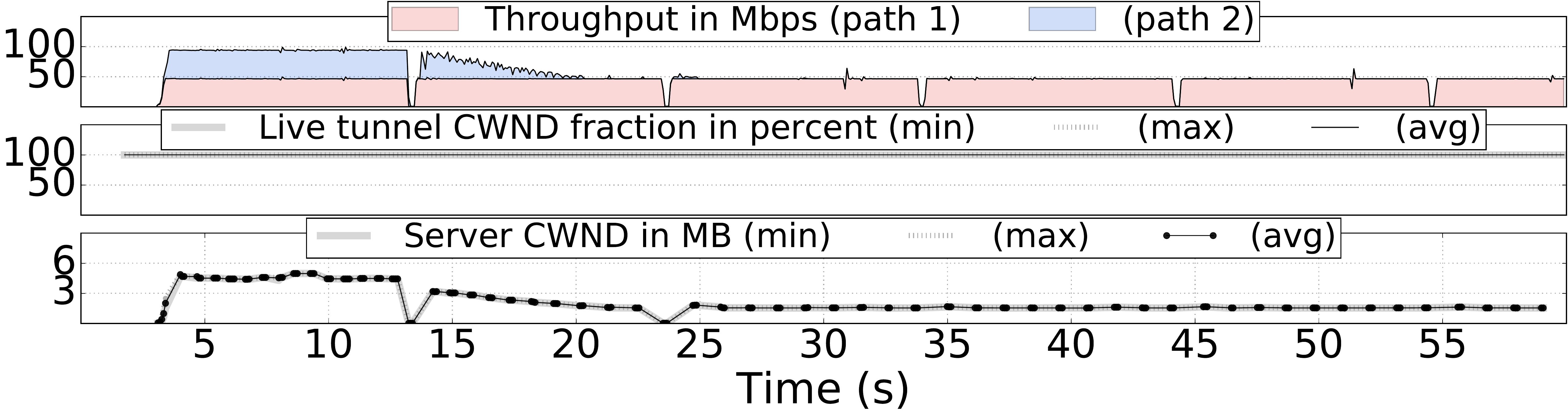}
\label{fig:ts-bbr-bbr-pp}}
\subfloat[][ACPF, short RTT, E2E CCA: TCP-BBR.]{
\includegraphics[width=.48\linewidth]
{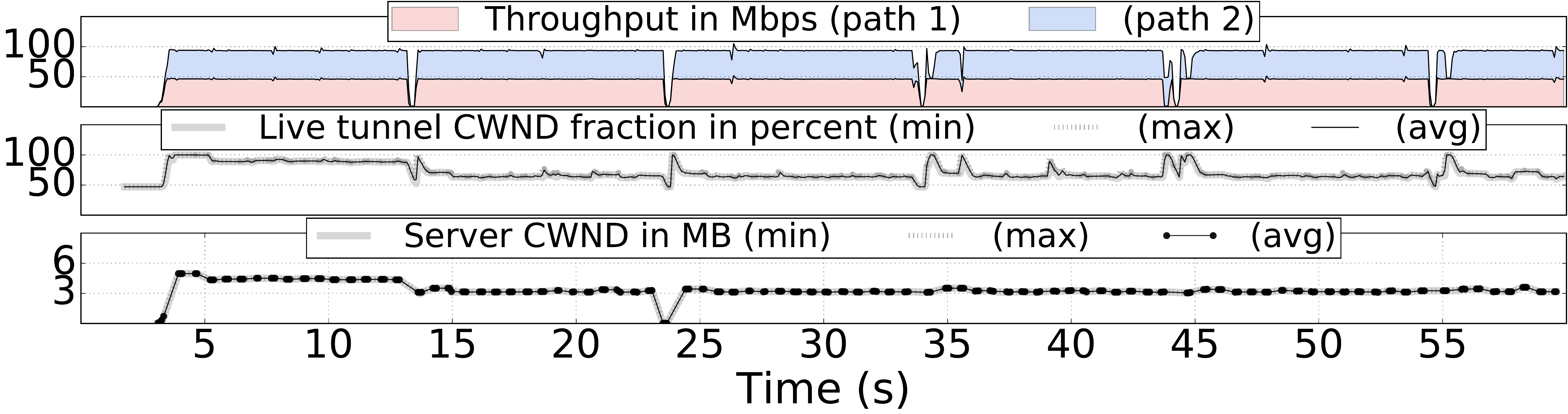}
\label{fig:ts-bbr-bbr-ppp}}

\caption{User-space comparison of CPF (left) \& ACPF (right) for different end-to-end CCAs.}
\label{fig:ts-multi-to-bbr-comp}
\end{figure*}

We present results from both the user and kernel-space frameworks. We leverage the former for its ability to also use TCP, its extensive inbuilt tracing, and the ease of running large sets of test. Using TCP-BBR as tunneling CCA is attractive since it is more mature than its DCCP counterpart.

\subsection{Evaluation setup}
\label{sec:imp-setup}

We used Mininet for our evaluation with the topology in Figure \ref{fig:topology}. In each case, values in parenthesis are for user-space tests. We set the bottleneck \textit{capacity} for both paths to 25 (50) Mbps. DCCP-CCID5 (TCP-BBR) is used as tunneling protocol. TC-TBF \cite{tc-tbf} is used to limit the capacity, and TC-Netem \cite{tc-netem} is used to emulate propagation delay. This setup allow us to vary the anchor deployment. As a default, we set a 4 ms \textit{latency} between anchor and server, and 16 and 26 ms over the primary and secondary path, respectively --- our \textit{short} scenario. We also use a \textit{long} RTT scenario where the aforementioned delays instead are 8 ms, 32 ms, and 42 ms. Both cases result in a path \textit{latency asymmetry} of 10 ms. The bottleneck buffer is large enough to not overflow from BBR. TCP-Iperf is used to generate a \textit{greedy} flow.

\subsection{Overview evaluation}
\label{sec:eval-main}

\begin{figure}
\centering
\includegraphics[width=.98\linewidth]
{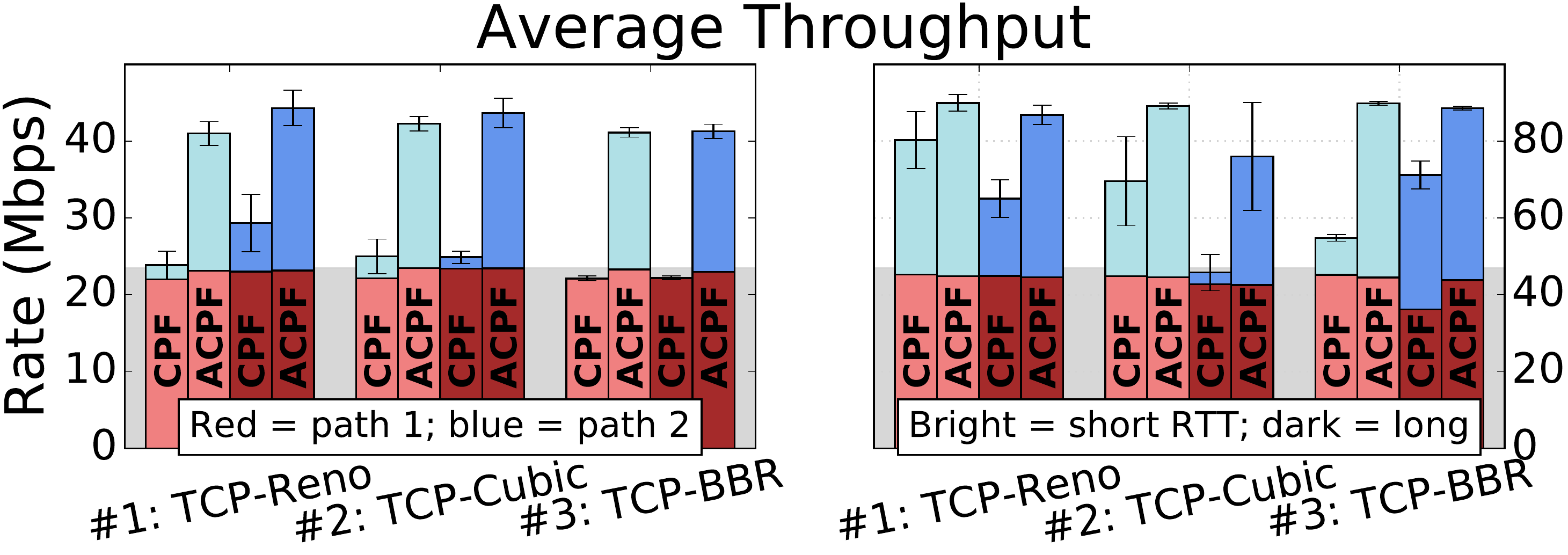}
\caption{Kernel-space (left) \& user-space (right).}
\label{fig:tp-bar-comp}
\end{figure}

Figures \ref{fig:ts-comp-kernel-cpf-short} to \ref{fig:ts-comp-kernel-acpf-long} show throughput obtained from the kernel-space framework. CCID5 is used as the tunnel CC in all cases, while TCP-NewReno, TCP-Cubic, or TCP-BBR is used end-to-end. The full set of tests is repeated for a short and long RTT. The path throughput is \textit{stacked}, and should ideally reach approximately the \textit{sum of both paths}. Each result was picked in a representative way from a set containing 5 repetitions, and illustrate that ACPF improves aggregation over CPF in \textit{each} scenario. The average throughput for each set is depicted in Figure \ref{fig:tp-bar-comp}, with an advantage to ACPF in all scenarios, ranging from 69\% - 86\% improvement in throughput. 



Figures \ref{fig:ts-reno-bbr-pp} to \ref{fig:ts-bbr-bbr-ppp} further illustrate why ACPF is able to achieve better aggregation than CPF. Each Figure depicts the result for a single \textit{user-space} test for \textit{3 parameters} over time:
\begin{enumerate*}[label=(\roman*)]
\item \textit{throughput};  
\item \textit{live CWND};  
\item \textit{server CWND}.
\end{enumerate*}
TCP-BBR is used in the tunnel in all cases, but the test is otherwise the same as previous. Figures \ref{fig:ts-reno-bbr-pp}, \ref{fig:ts-cubic-bbr-pp}, and \ref{fig:ts-bbr-bbr-pp}, show low aggregation for CPF. Figures \ref{fig:ts-reno-bbr-ppp}, \ref{fig:ts-cubic-bbr-ppp}, and \ref{fig:ts-bbr-bbr-ppp} show good aggregation for ACPF. These cases were \textit{carefully selected} from a set of 10 repetitions each and represent those repetitions where default CPF performs \textit{particularly bad}. Note that when the server uses TCP-NewReno or TCP-Cubic, the \textit{ProbeRTT} periods that occur every 10 seconds tend to be followed by a period of severe under-utilization of the secondary path. These periods clearly solicit an adverse reaction from the end-to-end CCA. One interesting case is shown in Figure \ref{fig:ts-reno-bbr-ppp} around the 35 second mark; the $CWND_p$ fraction crashed quite quickly to near its minimum value and $CWND_1$ \textit{would} have accepted most of $CWND_s$ and thereby prevented aggregation, \textit{had} the fraction been at 100\% after this event. Also, the user-space tunnel cause TCP-Cubic to leave slow-start early, often before utilization of the secondary path and sometimes at the very beginning of the flow. The latter is a particularly bad outcome and causes a high standard deviation. The average throughput for each set is again depicted in Figure \ref{fig:tp-bar-comp}, with an improvement of ACPF in all scenarios ranging from 24\% - 66\%. One encouraging result is that ACPF reliably yields nearly perfect aggregation for all scenarios, with the TCP-Cubic exception. 



\subsection{Detailed evaluation}
\label{sec:res-in-depth}

\begin{figure}
\centering

\subfloat[][Average throughput in Mbps.]{
\includegraphics[width=.98\linewidth]
{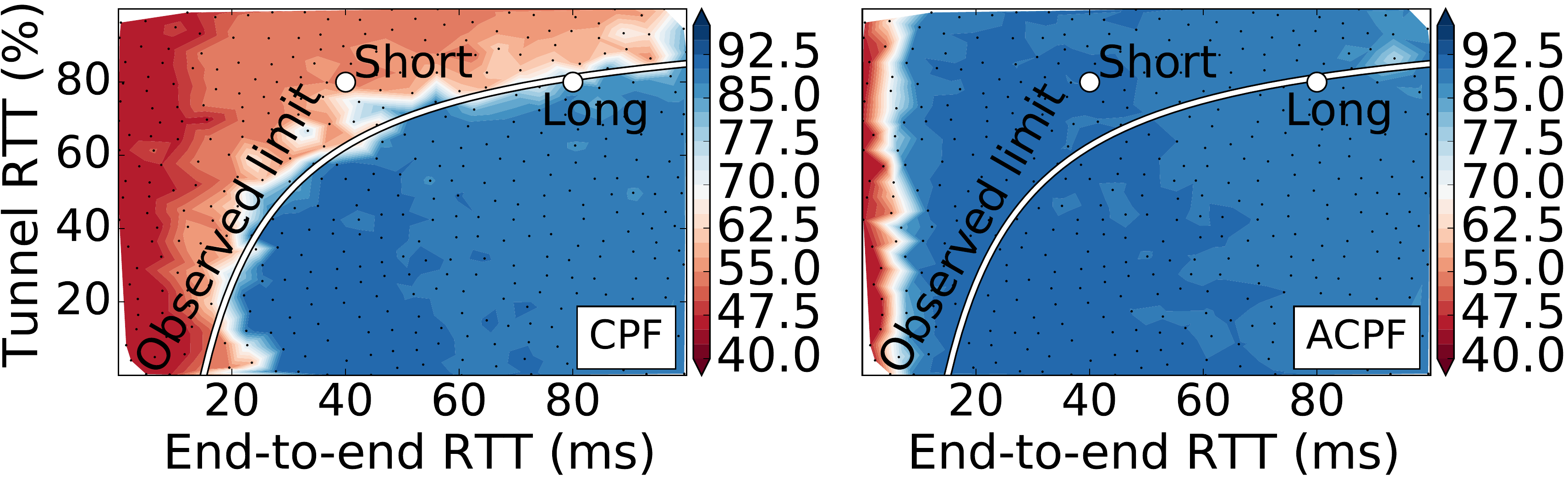}
\label{fig:cfpp-map-tp-comp}} \par

\subfloat[][Average live $CWND_p$ fraction in percent.]{
\includegraphics[width=.98\linewidth]
{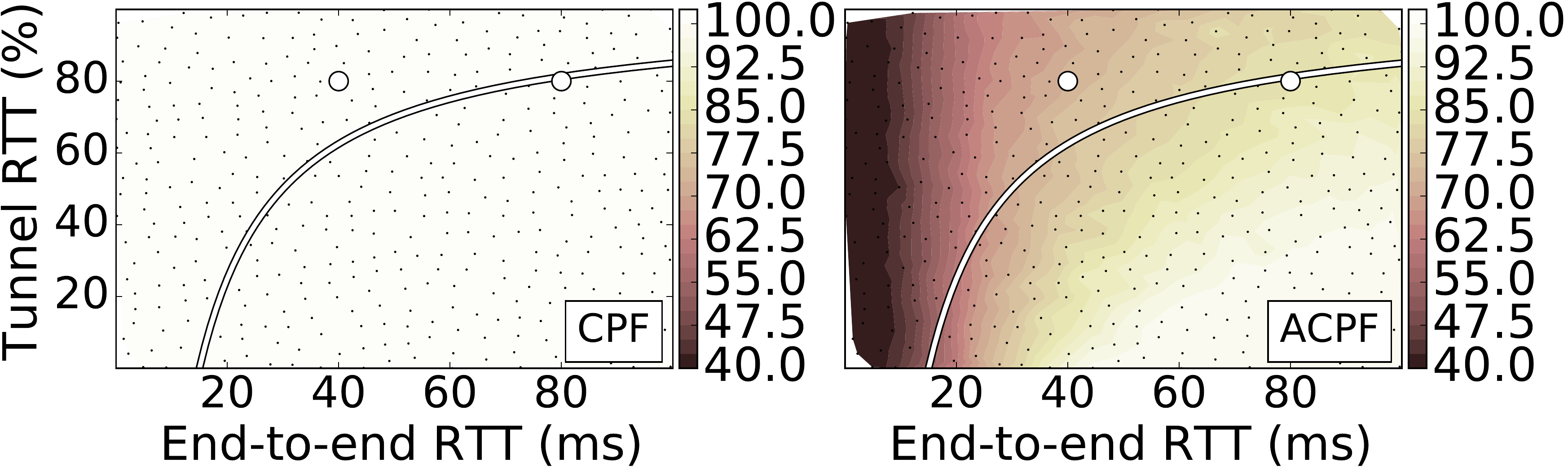}
\label{fig:cfpp-map-wf-comp}} \par

\subfloat[][Average server $CWND_s$ in MB.]{
\includegraphics[width=.98\linewidth]
{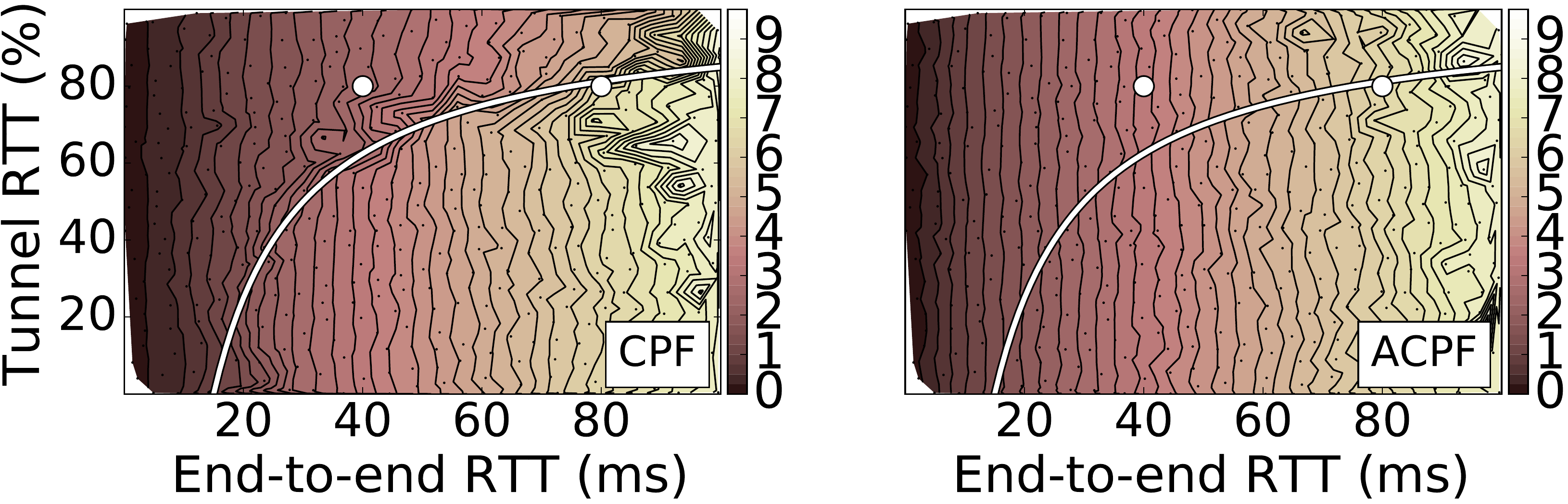}
\label{fig:cfpp-map-cw-comp}} \par

\subfloat[][Average server $CWND_s$ as a function of end-to-end RTT.]{
\includegraphics[width=.98\linewidth]
{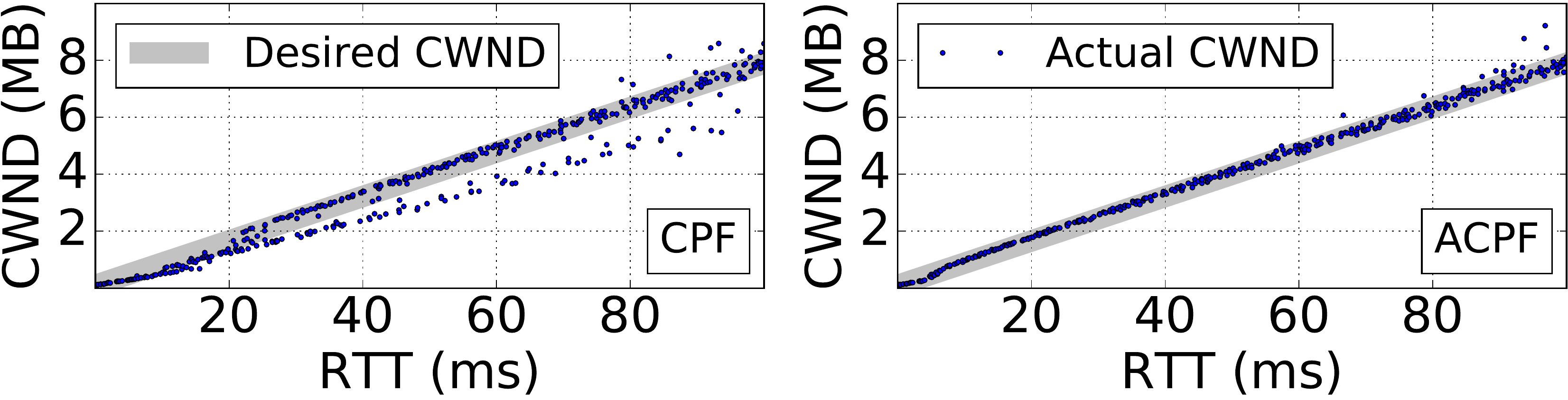}
\label{fig:cfpp-map-cw2-comp}}

\caption{Outcome for CPF (left) \& ACPF (right), for different RTT values \& anchor deployment locations.}
\label{fig:cfpp-map-comp}
\end{figure}

The CPF issue was first discovered when tunneling BRR-over-BBR in the user-space framework via the process described here. For this, 400 points in a 100-by-100 square area are generated randomly, with the exception that any point \textit{too close} to previously generated points is discarded and re-generated. The coordinates of these points are then translated into the end-to-end RTT in milliseconds, and the \textit{percent} of that RTT which is within the tunnel --- \textit{as opposed to between anchor and server} --- and are used as the configuration input for a test. The outcome is presented as a contour plot with each test marked, and interpolation between neighbouring tests.

Figure \ref{fig:cfpp-map-tp-comp} illustrates the throughput for all tests. A good outcome is deep blue, whereas a deep red outcome indicates \textit{no aggregation}. Both  
\begin{enumerate*}[label=(\roman*)]
\item a \textit{short end-to-end RTT}, and 
\item an \textit{anchor close to the server}
\end{enumerate*}
will inhibit aggregation for CPF. The reason for the elusive aggregation appears to be that $CWND_1$ is \textit{as large} as $CWND_s$. Since CPF will fill $CWND_1$ first, $CWND_s$ will have to be \textit{larger} than $CWND_1$ if any traffic is to "spill over" onto the expensive path. It is particularly difficult for the server BBR to overcome the tunnel BBR in this way, as BBR will have its $CWND$ size remain close to 2 BDP. Hence the clear patterns in Figure \ref{fig:cfpp-map-comp}.

Figure \ref{fig:cfpp-map-wf-comp} depicts the \textit{average} size of the live $CWND_1$ as a percent of the raw $CWND_1$. ACPF tends to have the \textit{smallest} live $CWND_1$ where CPF was \textit{least able to aggregate}, which illustrates that ACPF will not consistently deploy its offloading mechanism when it is not needed --- or alternatively, that ACPF functions more like CPF, when CPF is sufficient. 

Figure \ref{fig:cfpp-map-cw-comp} suggests that ACPF will trigger the server CC to \textit{perceive the second path}, and that the server therefore issues a \textit{larger} $CWND_s$. Note how the topographic lines \textit{veer to the right} above the white reference line for CPF. It \textit{should} be the case that BBR maintain $CWND_s$ as \textit{a function of the RTT}, but Figure \ref{fig:cfpp-map-cw2-comp} illustrates that this is \textit{only} the case for ACPF. The noise to the left in Figure \ref{fig:cfpp-map-cw2-comp} is caused by the server CCA sometimes \textit{losing track} of the secondary path, which causes BBR to perceive a different BDP --- \textit{resulting in the red regions in Figure \ref{fig:cfpp-map-tp-comp}}. A similar analysis may be done for other CCA combinations, but is omitted for lack of space.

\section{Conclusion and Future Work}
\label{sec:end-main}



In this paper, we demonstrated a weakness of CPF, the reason behind this flaw, and how to overcome it. The solution presented is a modified version named ACPF that is able to achieve path aggregation when default CPF is unable, and thereby increase throughput by 24\% - 86\% in the scenarios studied. ACPF enable the end-to-end CC to \textit{perceive} parallel paths \textit{more as a single path}. We used \textit{greedy} E2E flows to confirm that ACPF aggregates paths more or less as readily \textit{as if} there was a single path by limiting \textit{initial} usage of the cheapest path to 1 BDP. Whether aggregation is \textit{actually desirable} for any given greedy flow is arguably a decision made by e.g. the user or by a management agent.

There should be additional benefits to ACPF, e.g. an ability to perform a \textit{soft handover} from a degrading link without experiencing a temporary increase in delay. Of course, for such a switching scenario, ACPF may \textit{needlessly} increase the utilization of the expansive path if it \textit{underestimates} the BDP of the cheap path; a good BDP estimation is a prioritized goal. ACPF was developed in kernel-space MP-DCCP, as well as a user-space multipath framework. The former allow us to learn the BDP from the CC state, thus removing the need to \textit{infer} the BDP. It would also be desirable to be able \textit{contain bursts} on a single path to avoid using the secondary path too aggressively. These are left as future work.

\bibliographystyle{ACM-Reference-Format}
\bibliography{sample-base}

\end{document}